# Supply of engineering techniques and software design patterns in psychoanalysis and psychometrics sciences

Omid Shokrollahi

## Abstract:

The purpose of this study is to introduce software technologies and models and artificial intelligence algorithms to improve the weaknesses of CBT (Cognitive Behavior Therapy) method in psychotherapy. The presentation method for this purpose is the implementation of psychometric experiments in which the hidden human variables are inferred from the answers of tests. In this report, we describe the various models of Item Response Theory and measure the hidden components of ability and complementary parameters of the reality of the individual's situation. Psychometrics, selecting the appropriate model and estimating its parameters have been introduced and implemented using R language developed libraries. Due to the high flexibility of the Multi variant Rasch mixture Model, machine learning has been applied to this method of data modeling. BIC and CML were used to determine the number of hidden classes of the model and its parameters respectively, to obtain Measurement Invariance. The sensitivity of items to hidden attributes varies between groups (DIF), so methods for detecting it are introduced. This simulation is done based on the Verbal Aggression Dataset. We also analyze and compile a reference model based on this certificate based on the discovered patterns of software engineering. Other achievements of this study are related to providing a solution to explain the reengineering problems of the mind, by preparing an identity card for the clients by an ontology. Finally, applying the developed knowledge in the form of system thinking and recommended patterns in software engineering during the treatment process is pointed out.

## Introduction:

Cognitive Behavior Therapy (CBT) is a psychotherapeutic approach. In this type of psychotherapy performed by a specialist besides regular and purposeful tasks, a new thinking and feeling is formed in the brain and the person looks at the goals with a more acceptable attitude, and exchange more appropriate feedback with the his work and life. In this way, the system which is combination of the imaginative and the real world are examined. This dynamic system is described by variables.

Hidden traits (traits that cause visible behaviors) change over time. New hidden key variables appear and replaced. Examining them and their evolution is the main topic of psychotherapists today. Due to the limited capacity of the human brain, the achievements of complicated computer computation could be a dynamic approach to examining new variables and causal relationships that govern them, and address a person's observable behaviors and potentially emerging behaviors.

The main question: which models should the psychotherapist use to advance the treatment process and create a new look and behavior and make a lasting change?

Objective: introducing technologies and software models and artificial intelligence algorithms to improve the weaknesses of CBT (Cognitive Behavior Therapy) in psychotherapy.

## 1-1 Machine Learning Psychometrics:

In this phase, test information is analyzed. Psychometrics is the science of quantitatively studying psychological properties. Item Response Theory is a modern theory that measures these individual factors and differences, and the level of an individual's ability for answering a specific question at the scale being measured. To study the growth and inhibition of psychological variables over time and in a more comprehensive model, obtained information from various methods and evidence are integrated and according to IRT and various evidences and impact models, dynamic effect of variables from outside the system on these variables derived from psychological characteristics (which applying them could be done with the fuzzy techniques and conditional probability relations) and the progress of treatment and its feedback could be predicted and evaluated in defining the plans of each stage. Machine learning methods help to organize, analyze and compare and interpret this data in the form of modeling hidden features, abilities and tendencies. These computational tools are applied in ambiguous situations when dealing with file compromise. Practical concepts better explain the parameters of the problem and suggest reliable strategies for treatment and its scheduling.

## 1-2 Mind Reengineering Process:

The next step in providing a successful solution is to explain the reengineering problems of the mind from an ontology perspective. As we have said, to make a lasting difference, the psychotherapist manages the treatment plan with information from psychometrics based on specific patterns. Also, the effect and relationship of the individual's life system (family, work environment, community) along with these variables could be examined.

These patterns are introduced into the migration strategy and in this study, evidence is provided that the recommended patterns in software engineering help to modify and re-engineer mental models and stabilize the changes.

Finally, a strategy for using intelligent mental structure design is presented to help people correct errors in their mental structure. Those solutions are applied in an acceptable way in process of systemic thinking and modeling according to the situation, abilities and hidden attributes.

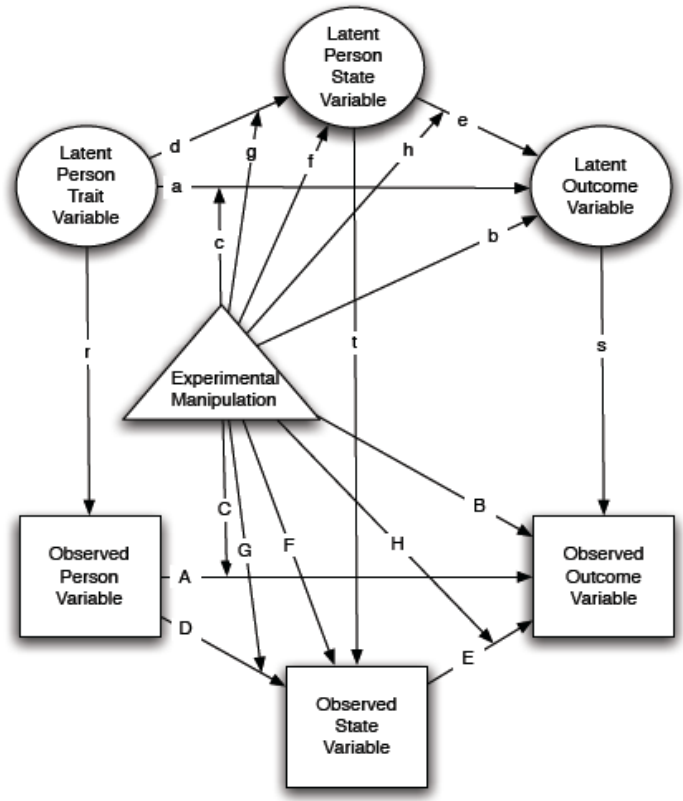

**Fig. 3.1** Both experimental and observational research attempts to make inferences about unobserved latent variables (traits, states, and outcomes) in terms of the pattern of correlations between observed and manipulated variables. The uppercase letters (A-F) represent observed correlations, the lower case letters (a-f) represent the unobserved but inferred relationships. The shape of the mappings from latent to observed (r, s, t) affect the kinds of inferences that can be made(Adapted from Revelle (2007) )

And we want to find the hidden structures in the best way from the observed quantity. The developed libraries in R which are completed in collaboration with psychotherapy teams, provide a framework for" reproducible statistical analysis" that after a variety of multivariate statistical analysis and dimension reductions, which are easily implemented with the functions of these libraries for estimation, organization and use in structural equation models.

## 2-1-1- Implementation of Item Response Theory:

Reliability in modern psychometrics is not only the result of items but also of the data source. IRTs is the concept of reliability has been developed in this way. The reasons for using these models is that IRTs are not dependent test and sample which are dichotomous or polytomous which could be used as a basis for Computerized Adaptive Testing.

IRTs consider a variety of parameters and provide reliable measure as a suitable tool for describing and predicting the behavior of hidden factors. [ltm package] a library in R, is developed to handle when modeling a test hypothesis depends on several attributes (Multi Dimensional).

More advanced models from the Rasch family have also been explored. [extended rasch models] [polytomous multiple latent variables] in the R language, functions have been developed to use different types of IRTs. To complete these approaches, we must also refer to uni- and multidimensional IRT modeling, nonparametric IRT modeling, and Differential Item Functioning. Additional heterogeneity can also be modeled in their form (such as mixture models). [] MultiCIRT library functions perform the parameterization of these models. Test Equating facilitates multi-group matching.

With the plink () method, one-dimensional models that even have hybrid parameters could be valued and equalized.

It is noteworthy that in modeling hidden attributes, the structure of relationships influences the introduction of other models that have validity and can affect the overall picture.

Rasch Model is a set of IRT models that formulate the function of probability that it's response is correct in terms of ability. Depending on the sensitivity, in the simplest form, a logistics model with 2 to 4 parameters is made that each of model have a specific behavioral effect [].In the following formulas, you can see the general form of this family.

"Mixture Model" is a generic approach to implement this assumption that the observed data is rooted in the known probability of different models. The following formula has been used to determine the parameters of Mixture Model that Conditional Maximum Likelihood Estimation algorithms. [psychomix]

$$k = 1, \ldots, K:$$
$$(\hat{\beta}^{(k)}, \hat{\delta}^{(k)}) = \underset{\beta^{(k)}, \delta^{(k)}}{\operatorname{argmax}} \sum_{i=1}^{n} \hat{p}_{ik} \log f(y_i | \beta^{(k)}, \delta^{(k)})$$
$$= \left\{ \underset{\beta^{(k)}}{\operatorname{argmax}} \sum_{i=1}^{n} \hat{p}_{ik} \log h(y_i | r_i, \beta^{(k)}); \underset{\delta^{(k)}}{\operatorname{argmax}} \sum_{i=1}^{n} \hat{p}_{ik} \log g(r_i | \delta^{(k)}) \right\}.$$

Test_dim () and class_item () are two functions in this library to determine the optimal number of model dimensions which use the Hierarchical clustering and Nested Model methods to implement this calculation. (Matching the observed score with the actual score) are performed by the kernel method. ( vonDaviier, The Netherlands )

## 2-1-2 Differential Item Functioning:

Regardless of the bias differentiation, there are differences between different groups in the statistical community which misinterpretation of the difference in scores leads to a wrong hypothesis.

Consider a questionnaire that measures the depression difference between the sexes. The item "How many times have I felt depression in the past week?" Or "I felt hopeless about the future last week" is expressed almost equally between men and women. But in the item: "Last week I cried easily or felt like crying" has a much higher threshold among men than women( schaeffer 1988; Steinberg (2006)). What was measured as difference in item difficulty may have different origins. For example, the scatter of differences in the sensitivity degree of an item to a hidden attribute was presented.

These R packages are tools for this purpose.

[new method for detecting differential item functioning in the Rasch]

[Differential Item Functioning using IRT]

[R Package for detecting differntal item functioning]

## 2-1-3 Simulated problem:

Our aim is providing a model in which the parameters of the latent classes explaining each individual's ability which are independent of measurement and statistical distribution of ability and of individuals response so that the estimation of individual attribute parameters could be measured purely in the experiment. (measurement invariance)

We do this implementation on Verbal Aggression data [(De Boeck and Wilson 2004)].This hypothesis that the school type affects the test result has been tested. The data is modeled with Rasch Mixture. In this method of modeling, there is measurement invariance but this hypothesis is violated in collecting data that recognize the unobserved heterogeneity in response of data to items. With this model, we also recognize intergroup differentiation (DIF).

General DIF diagnostic methods such as the LR test and the Rasch tree can make this distinction. [Strobl et al. 2014, Ankenmann, Witt, and Dunbar 1999] These methods are done based on the conditional part of validation. Adding an additional parameter called the score distribution can also justify measurement invariance.

The BIC method is suitable for selecting the number of IRT classes of two-choice models. "Conditional Maximum Likelihood" is used repeatedly to estimate the parameters of this model until convergence is achieved.

$$L(\pi^{(1)},\ldots,\pi^{(K)},\beta^{(1)},\ldots,\beta^{(K)},\delta^{(1)},\ldots,\delta^{(K)}) = \prod_{i=1}^{n}\sum_{k=1}^{K}\pi^{(k)}f(y_i|\beta^{(k)},\delta^{(k)})$$
$$= \prod_{i=1}^{n}\sum_{k=1}^{K}\pi^{(k)}h(y_i|r_i,\beta^{(k)})\,g(r_i|\delta^{(k)}). \quad (4)$$

Model parameters are determined by Iterative Hybrid Ordinal Logistic Regression, Item Response Theory techniques. The corresponding R code is located in the codes folder.

Participants face two difficult situations, and their verbal response to violence (cursing, eyebrow-raising, and shouting) is assessed in 12 ways.

**S1WantCurse S1DoCurse S1WantScold S1DoScold S1WantShout S1DoShout**

**S2WantCurse S2DoCurse S2WantScold S2DoScold S2WantShout S2DoShout**

## 2-2 Reengineering of the mind

By recognizing the root of many psychological problems, the errors of hierarchical relationships are understood beside. Consequences of using defective operators by using subjective knowledge, depending on the nature in main focus of the research at this stage, is the use of more advanced technology to build a complete " model reference " to solve problems and find solutions. This explanation should include a complete list of symptoms and consequences of the problem, and an accurate description of the cause roots, the used method, the forces and pressures whose balance has been disturbed or neglected or misused be distinguished . As a scientific researcher, when introducing the program, in the form of a model, we must formally present our achievement in the form of answers to these questions.

Some of a person's psychological problems are defective and inappropriate integration of mental classes (a concept in the object-oriented design approach), violation of encapsulation, and so on. To deal with these problems, new scientific cognitive-therapeutic methods must be introduced. There is ontology to explain the migration strategy in the form of structure for this process. Attempts to identify and prepare the mind must be made within a scientific framework of specific considerations and at the same time for designing multiple tests. These challenges must be addressed in practice for each case. One of the challenges of CBT is the uncontrolled growth of latent variables that may emerge over time or elsewhere. In the next phase, we seek to answer the question about the effect and relationship of the individual's life system (family, work environment, community) with these variables.

The three software engineering concepts "Patterns, Antipatterns, and Refactoring" are used here. In each example of the proposed design, the recommended patterns (with proven efficiencies) are inspected for avoiding of antiplatelet effects .

Patterns are predicted at all three levels: 1. The level at which the tested person is solving the problem and executing the plans. 2- The level at which the system architected and the process of structuring is implemented, and 3- The level at which the process is progressed and guided. At all levels, these processes must be properly managed.

Each designer must answer these questions in an ID to justify the pattern.

First, what problem or problems should be solved and in which context it could be used?

Provide diagrams and descriptions of the problem and solution.

Introduce structural and dynamic diagrams and interactions.

Provide an implementation guide.

Which names or ontologies could be used to introduce it?

Involves both parts (of the brain) and objects (intrinsically formed concepts).

What is the function and internal relationship between these components? The template is used for creating Creational, Behavioral , and Structural structure. What are the consequences, strengths, and weaknesses of this method? What is gained and lost in their use, and what are their known applications?

| Scenario | Latent class I | | Latent class II | |
|---|---|---|---|---|
| | Mean abilities | Difficulties | Mean abilities | Difficulties |
| *No impact* ($\Theta = 0$) | | | | |
| 1   no DIF ($\Delta = 0$) | $\{0\}$ | $\beta^I$ | — | — |
| 2   DIF ($\Delta > 0$) | $\{0\}$ | $\beta^I$ | $\{0\}$ | $\beta^{II}$ |
| *Impact* ($\Theta > 0$) | | | | |
| 3   no DIF ($\Delta = 0$) | $\{-\Theta/2, +\Theta/2\}$ | $\beta^I$ | — | — |
| 4   DIF ($\Delta > 0$), not coinciding | $\{-\Theta/2, +\Theta/2\}$ | $\beta^I$ | $\{-\Theta/2, +\Theta/2\}$ | $\beta^{II}$ |
| 5   DIF ($\Delta > 0$), coinciding | $\{-\Theta/2\}$ | $\beta^I$ | $\{+\Theta/2\}$ | $\beta^{II}$ |

Table 1: Simulation design. The latent-class-specific item parameters $\beta^I$ and $\beta^{II}$ differ by $\Delta$ for two elements and thus coincide for $\Delta = 0$, leaving only a single latent class.

Estimation of item parameters depends on the distribution of scores in the Rasch Mixture Model. Two distributions are shown in Figure .. (saturated and mean-variance specification). In Figure ... dataset result representation and the questionnaire results are shown.

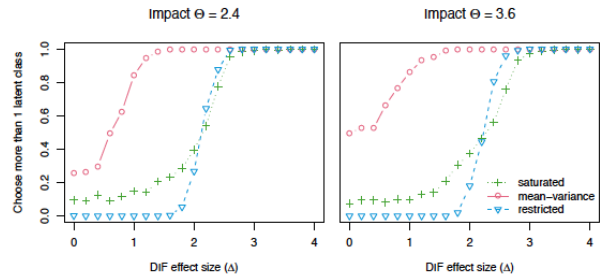

Figure : Rate of choosing a model with $\hat{K} > 1$ latent classes for data from Scenario 5 (impact and DIF, coinciding).

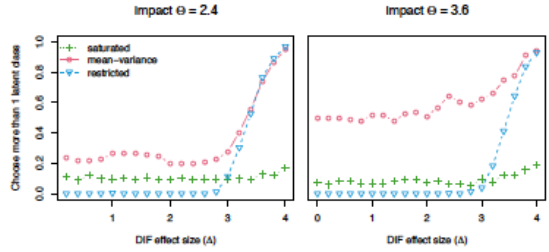

Figure : Rate of choosing a model with $\hat{K} > 1$ latent classes for data from Scenario 4 (impact and DIF, not coinciding).

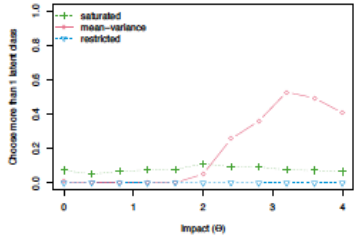

Figure : Rate of choosing a model with $\hat{K} > 1$ latent classes for data from Scenario 3 (impact without DIF, i.e., $\Delta = 0$).

The simple RASCH model is not suitable and more than one scale is required for measurement. Tables 2 and 3 show how we can obtaine the number of hidden classes K = 3. This number explains the distribution of points for each scenario.

The evaluation results of each model are displayed in " model assessment" image and the difference between the samples is introduced with the additional variable "score distribution" and thus "Rasch Mixture modeling", and the image results (considering score distribution) ensures that with this structure ,estimation of parameters is independent of the ability distribution and is sensitive to the hidden structure in the difficulty of item.

| Model | k | #Df | log L | BIC |
|---|---|---|---|---|
| restricted (mean-variance) | 1 | 13 | −1900.9 | 3874.6 |
| restricted (mean-variance) | 2 | 25 | −1853.8 | 3847.8 |
| **restricted (mean-variance)** | **3** | **37** | **−1816.9** | **3841.4** |
| restricted (mean-variance) | 4 | 49 | −1792.0 | 3858.8 |

Table 2: DIF detection by selecting the number of latent classes $\hat{K}$ using the restricted Rasch mixture model.

| Model | k | #Df | log L | BIC |
|---|---|---|---|---|
| saturated | 3 | 65 | −1795.2 | 3955.1 |
| restricted (saturated) | 3 | 45 | −1814.1 | 3880.6 |
| mean-variance | 3 | 41 | −1812.2 | 3854.4 |
| **restricted (mean-variance)** | **3** | **37** | **−1816.9** | **3841.4** |

Table 3: Selection of the score distribution given the number of latent classes $\hat{K} = 3$.

# 4- Conclusion:

Sophisticated treatment processes in the cognitive behavioral therapy approach require new solutions to improve the interaction between the psychotherapist and the client and reduce the unwanted costs of treatment. In this process, two components such as Mathematical detection and modeling are of fundamental importance. The measured psychological variables with sufficient complexity must explain the hidden variables correctly.

First, it introduces the strengths of using machine learning-based models, then, to estimate the hidden abilities, traits, and attributes of the person being tested by relying on Item Response Theory, a number of statistical models for psychometrics are introduced and tested.

Item variable response model and latent variable response model were examined on the Verbal aggression database

Item variable response and latent variable response model were examined on the Verbal aggression database.

We further showed that to achieve the right attitude and reconstruct and re-engineer the minds of software knowledge systems , design knowledge could be a more appropriate approach to problems and provide effective insights for this purpose to the psychotherapist to create a new look and behavior and stable change. First, the process of designing an identity card for each case was introduced. Patterns and Anti patterns and refactoring software engineering concepts were introduced to advance the treatment process based on this certificate. These patterns are introduced in the form of migration strategy and in this research, evidence is provided that the patterns recommended in software engineering help to modify and re-engineer mental models and stabilize the changes provided that those solutions are accepted in an acceptable way, with systemic thinking and systematic modeling appropriate to the situation, ability and hidden attributes of the subject. The migration strategy provides a strategy for using intelligent mental structure design to help people correct errors in their mental structure.